\documentclass[
    ,final            
  ]
  {aipproc}

\layoutstyle{6x9}
\newcommand{\be}{\begin{equation}}\newcommand{\ee}{\end{equation}}
\newcommand{\bea}{\begin{eqnarray} }\newcommand{\eea}{\end{eqnarray}}
\newcommand{\beaa}{\begin{eqnarray}}\newcommand{\eeaa}{\end{eqnarray}}
\newcommand{\ba}{\begin{array}}\newcommand{\ea}{\end{array}}
\newcommand{\bit}{\begin{itemize}}\newcommand{\eit}{\end{itemize}}
\newcommand{\ben}{\begin{enumerate}}\newcommand{\een}{\end{enumerate}}

\def\lf{\left}

\def\pa{\partial}\def\ran{\rangle}\def\rar{\rightarrow}
\def\ri{\right}

\def\al{\alpha}\def\bt{\beta}\def\ga{\gamma}

\def\te{\theta}
\def\si{\sigma}

\def\lf{\left}

\def\pa{\partial}\def\ran{\rangle}
\def\rar{\rightarrow}
\def\ri{\right}

\def\al{\alpha}\def\bt{\beta}\def\ga{\gamma}

\def\te{\theta}

\def\si{\sigma}

\def\1{{_{1}}}\def\2{{_{2}}}
\def\nof{:\;\!\!\;\!\!:}
\begin{document}

\title{Flavor states of mixed neutrinos}


\author{M. Blasone}{
  address={Istituto Nazionale di Fisica Nucleare (INFN), Gruppo Collegato
di Salerno and DMI,
 Universit\`a di Salerno,
 Fisciano (SA) - 84084 Italy}
}

\author{A. Capolupo}{
  address={Department of Physics and Astronomy,
University of Leeds, Leeds LS2 9JT UK}
}

\author{C. R. Ji}{
 address={ Department of Physics, North Carolina State
University, Raleigh, NC 27695-8202, USA}
}

\author{G. Vitiello}{
  address={Istituto Nazionale di Fisica Nucleare (INFN), Gruppo Collegato
di Salerno and DMI,
 Universit\`a di Salerno,
 Fisciano (SA) - 84084 Italy}
}

\begin{abstract}

By resorting to previous results on flavor mixing in
Quantum Field Theory,
we show how to consistently define flavor states of mixed neutrinos
as eigenstates of the flavor charge operators.
 \end{abstract}

\maketitle

The issue of a proper definition of flavor states for mixed neutrinos
has been object of discussion in recent years \cite{BV95}-\cite{Blasone:2006jx}: although
the usual
Pontecorvo states \cite{Bilenky:1978nj,Kimbook}
represent a valid tool for describing
the main physical features of neutrino oscillations,
it has been clear since some time  that conceptual
problems arise in connection with a proper definition of flavor
states. In fact, it was even stated \cite{Kimbook} that it is
impossible to construct such states and a formalism has been
developed with the aim to avoid their use in the calculation of
oscillation probabilities \cite{Beuthe:2001rc}.

The root of such difficulties has been found by studying flavor mixing
at level of quantum fields. It
has then emerged \cite{BV95} that the vacuum for neutrino fields with definite
masses turns out to be unitarily inequivalent to
the vacuum for the flavor neutrino fields. The
condensate structure of the flavor vacuum leads to a modification of flavor
oscillation formulas \cite{BHV98,binger,JM03,yBCV02},
exhibiting new features with respect to the usual quantum mechanical
ones \cite{Beuthe:2001rc,Giunti:2007ry}. Further developments
include Lorentz invariance violation \cite{nulorentz}
and neutrino mixing contribution to the dark energy of the Universe
\cite{pascos2}.

In the Standard Model, flavor neutrinos are produced in charged current
weak interaction processes, like $ W^{+} \rightarrow e^{+} + \nu_{e}$. At tree level,
flavor charge is strictly conserved in the vertex;
the presence of mixing allows for the possibility of violation of lepton
number via loop corrections. Such corrections are however extremely
small and practically unobservable. Therefore one can assume the tree level
as a good approximation of the real processes in which neutrinos are created.
This is what is done in practice: neutrinos are identified by the observation
of the corresponding charged leptons, {\it assuming} flavor conservation in
production/detection vertices.
It is therefore essential to have a clear understanding
of the neutrino flavor charges and states in the presence of mixing.

One could nevertheless insist in {\it declaring} that
flavor states are not to be used because everything can be written
in terms of neutrino fields with definite masses and consequently
solely neutrino mass
eigenstates are being used.
This is, however, like  ``to sweep the dust under the carpet'', since
in this way one operates a selection in favor of the Hilbert space of
the mass eigenstates.
Such a choice has no mathematical basis, because of the existence of a
separate Hilbert space (the one for the flavor eigenstates). It is also
lacking of physical motivations: indeed, in practice, it adopts the Pontecorvo states which
are not eigenstates of the flavor charges, in contrast with the observed
conservation of lepton number in the neutrino production/detection vertices.

We therefore devote the present report to: a) properly define flavor
charges and states and b) to estimate how much lepton charge is
violated by Pontecorvo states.

\vspace{0.5cm}

Let us start by  considering  the charges
for flavor (mixed) neutrinos \cite{currents,Blasone:2006jx}.
We discuss here the case of mixing between two generations.
Extension to three generations and beyond can also be done
\cite{JM02,yBCV02}.

For our purposes, the relevant Lagrangian density terms are
  \bea\label{lagrangian2}
  {\cal L} & =
& \, {\cal L}_{0}  +{\cal L}_{int}\,,
\eea
where ${\cal L}_{0}$
is the free lepton Lagrangian:
\bea\label{Leptlagrangian2} {\cal
L}_{0} \, = \, \lf({\bar \nu_e}, {\bar \nu_\mu}\ri) \lf( i \ga_\mu
\pa^\mu - M_{\nu} \ri)  \lf(\ba{c}\nu_e\\ \nu_\mu\ea\ri)\, + \,
\lf({\bar e},
{\bar \mu}\ri) \lf( i \ga_\mu \pa^\mu - M_{l} \ri)  \lf(\ba{c}e\\
\mu\ea\ri)\,, \eea
including the neutrino non-diagonal mass matrix
$M_{\nu}$ and the mass matrix of charged leptons $M_{l}$:
\bea &&
M_{\nu}\,=\,  \lf(\ba{cc}m_{\nu_e} & m_{\nu_{e\mu}}
\\ m_{\nu_{e\mu}} & m_{\nu_\mu}\ea\ri)    \quad ;\qquad
M_l\,=\,  \lf(\ba{cc}m_e &0 \\
0 & m_\mu\ea\ri)\, .\eea
${\cal L}_{int}$ is the charged current weak interaction
Lagrangian:
\bea\label{L-interact}  {\cal L}_{int} \, = \,
\frac{g}{2\sqrt{2}} \lf [ W_{\mu}^{+}(x)\,
\overline{\nu}_{e}(x)\,\gamma^{\mu}\,(1-\gamma^{5})\,e(x) +
W_{\mu}^{+}(x)\,
\overline{\nu}_{\mu}(x)\,\gamma^{\mu}\,(1-\gamma^{5})\,\mu(x) + h.c.
\ri]. \eea

Of course,  ${\cal L}_{0} $ can be diagonalized in terms of
neutrino fields $\nu_1$, $\nu_2$, with definite masses $m_1$, $m_2$. Here, however,
we are interested in the construction of flavor charges and therefore we consider
the above Lagrangian in terms of flavor fields. To this
end, we observe that ${\cal L}$ is invariant under the global phase transformations
\bea \label{ephase}  e(x) \rightarrow   e^{i \alpha}e(x) \,, \qquad
\nu_{e}(x) \rightarrow  e^{i \alpha} \nu_{e} (x)\,, \eea together
with \bea \label{muphase}
 \mu(x) \rightarrow   e^{i
\alpha}\mu(x) \,, \qquad \nu_{\mu}(x) \rightarrow   e^{i \alpha}
\nu_{\mu} (x)\,.
  \eea
These are generated by
\bea\label{Qflav} Q_{e}(t)& = & \int d^{3}{\bf x} \, e^{\dag}(x)e(x)
\,, \qquad Q_{\nu_{e}} (t) =  \int d^{3}{\bf x} \,
\nu_{e}^{\dag}(x)\nu_{e}(x)\,,
\\ [2mm] \label{QflavLept}
Q_{\mu}(t) & = &  \int d^{3}{\bf x} \,
 \mu^{\dag}(x) \mu(x)\,, \qquad Q_{\nu_{\mu}}(t) = \int d^{3}{\bf
x} \, \nu_{\mu}^{\dag}(x) \nu_{\mu}(x) \,, \eea
respectively. The invariance of the Lagrangian is then expressed by
\bea [Q_l^{tot}\,, \, {\cal L}] = 0\,, \eea
which guarantees the conservation of the total lepton number. Here,
$Q_l^{tot}$ is the total Noether (flavor) charge:
\be Q_{l}^{tot} = Q_{\nu_{e}}(t) + Q_{\nu_{\mu}}(t) + Q_{e}(t) +
Q_{\mu}(t) \,=\, Q_e^{tot}(t) \,+ \,Q_\mu^{tot}(t)\,, \ee
\bea \label{leptcharges}
Q_{e}^{tot} (t) = Q_{\nu_{e}}(t) + Q_{e}(t)\,, \qquad
Q_{\mu}^{tot}(t) = Q_{\nu_{\mu}}(t) + Q_{\mu}(t)\,.
 \eea
Note that the presence of the
mixed neutrino mass term, i.e.  the non-diagonal mass matrix
$M_\nu$,  prevents the invariance of ${\cal L}$
under the separate phase transformations (\ref{ephase}) and
(\ref{muphase}). Consequently, the flavor charges of Eq.(\ref{leptcharges}) are time
dependent.

However, family lepton numbers are still good quantum numbers if
the neutrino
production/detection vertex can be localized within a region much
smaller than the region where flavor oscillations take place.
This is what happens in practice, since
tipically the spatial extension of the neutrino source/detector is
much smaller than the neutrino oscillation length.

We thus look for the  flavor neutrino states as eigenstates of the neutrino
flavor charges
$Q_{\nu_e}$ and $ Q_{\nu_\mu}$. These operators may be
expressed in terms
of the (conserved) charges for the neutrinos with definite masses $Q_{\nu_1}$ and
$Q_{\nu_2}$ in the following way \cite{Blasone:2006jx}:
\bea\label{carichemix1}
&& \hspace{-1cm}Q_{\nu_e}(t) = \cos^2\te\;  Q_{\nu_1} +
\sin^2\te \; Q_{\nu_2} + \sin\te\cos\te \int d^3{\bf x}
\lf[\nu_1^\dag (x) \nu_2(x) + \nu_2^\dag(x) \nu_1(x)\ri]\,,
\\
\label{carichemix2}
&& \hspace{-1cm} Q_{\nu_\mu}(t)= \sin^{2}\te \; Q_{\nu_1}
+\cos^{2}\te \; Q_{\nu_2} - \sin\te \cos\te \int d^3{\bf x}
\lf[\nu_1^\dag(x) \nu_2(x) + \nu_2^\dag(x) \nu_1(x)\ri]\,. \eea
Notice that the last term in the above expressions
forbids the construction of eigenstates of
the $Q_{\nu_\sigma}(t)$, $\sigma=e,\mu$, in the Hilbert space
 ${\cal H}_{1,2}$ for the fields with definite masses.
Indeed the vacuum $|0\ran_{e,\mu}$ for the mixed field operators $\nu_\si$ turns out to be
orthogonal to the vacuum state $|0\ran_{1,2}$ for the fields with definite masses
$\nu_j$, with $j=1,2$ \cite{BV95}. One can also show that the above neutrino flavor
charge operators are diagonal in the ladder operators $\al_{\si}$, $ \beta_{\si}$, for the
neutrino flavor fields:
\bea\label{flavchadiag}
&&{}\hspace{-1cm}
\nof Q_{\nu_\sigma}(t) \nof
\; \,= \,\; \sum_{r}
\int d^3 {\bf k} \, \lf( \al^{r\dag}_{{\bf k},\si}(t) \al^{r}_{{\bf
k},\si}(t)\, -\, \beta^{r\dag}_{-{\bf k},\si}(t) \beta^{r}_{-{\bf
k},\si}(t)\ri)\, ,\qquad \sigma=e,\mu \eea
where $\nof ... \nof\,$ denotes normal ordering
with respect to  $|0\ran_{e,\mu}$.
This makes straightforward the definition (at reference time $t=0$)
of  flavor  neutrino and antineutrino states as:
\bea\label{flavstate} |\nu^{r}_{\;{\bf k},\si}\ran \equiv
\al^{r\dag}_{{\bf k},{\sigma}} |0\ran_{{e,\mu}}\quad; \qquad
|{\bar \nu}^{r}_{\;{\bf k},\si}\ran \equiv
\bt^{r\dag}_{{\bf k},{\sigma}} |0\ran_{{e,\mu}}\, ,\qquad \qquad\si =
e,\mu . \eea
which are thus by construction eigenstates of the operators  in Eq.(\ref{flavchadiag})
at $t=0$.

\vspace{.5cm}

Let us now turn to the quantum mechanical (Pontecorvo) flavor neutrino
states \cite{Bilenky:1978nj}:
\begin{eqnarray} \label{nue0a}
|\nu^{r}_{\;{\bf k},e}\rangle_P &=& \cos\theta\;|\nu^{r}_{\;{\bf k},1}\rangle \;+\;
\sin\theta\; |\nu^{r}_{\;{\bf k},2}\rangle \,,
\\ [2mm] \label{nue0b}
|\nu^{r}_{\;{\bf k},\mu}\rangle_P &=& -\sin\theta\;|\nu^{r}_{\;{\bf k},1}\rangle \;+\;
\cos\theta\; |\nu^{r}_{\;{\bf k},2}\rangle \, .
\end{eqnarray}
They  are
clearly {\em not} eigenstates of the flavor charges
\cite{Blasone:2005ae} as can be explicitly seen by using Eqs.(\ref{carichemix1})
and (\ref{carichemix2}).
In order to estimate how much the flavor charge is violated in the
usual quantum mechanical states, one can take the expectation values
of the flavor charges on the above Pontecorvo states.
Considering for example an electron neutrino
state, we obtain:
\bea\label{caric}  \;_{P}\langle\nu^{r}_{\,{\bf k},e}|:
Q_{\nu_e}: |\nu^{r}_{\,{\bf k},e}\rangle_{P} &=& \cos ^{4}\theta
+ \sin^{4}\theta + 2 |U_{\bf k}| \sin^{2}\theta \cos^{2}\theta \,<\,
1,
\\ [2mm]
\;{}_{P}\langle\nu^{r}_{\,{\bf k},e}|
:Q_{\nu_\mu}: |\nu^{r}_{\,{\bf k},e}\rangle_{P} &=& 2 \,(1 - |U_{\bf k}|)\,
\sin^{2}\theta \cos^{2}\theta  \,> \,0,
\eea
for any $ \theta \neq 0,\;  m_{1} \neq m_{2},\; {\bf k} \neq
0$ and
where  $: ... :\,$ denotes normal ordering with respect to the
vacuum state $|0\ran_{1,2}$.
In the relativistic limit, $|U_{\bf k}| \rar 1$ \cite{BV95,pascos2} and the Pontecorvo
states are a good approximation for the exact charge eigenstates
Eq.(\ref{flavstate}).
%
%
%


\medskip

We thank Arttu Rajantie and the organizers of PASCOS-07 for giving
us the opportunity to present this work at the PASCOS conference.
We also acknowledge INFN for financial support.

\end{document}